\documentclass[superscriptaddress,twocolumn]{revtex4}
\usepackage{graphicx}
\usepackage{amsmath}
\usepackage{color}

\begin{document}

\title{Fibre-optic metadevice for all-optical signal modulation based on coherent absorption}

\author{Angelos Xomalis}
\email{ax1c15@soton.ac.uk}
\affiliation{Optoelectronics Research Centre, University of Southampton, Southampton, SO17 1BJ, UK}
\affiliation{Centre for Photonic Metamaterials, University of Southampton, Southampton, SO17 1BJ, UK}

\author{Iosif Demirtzioglou}
\affiliation{Optoelectronics Research Centre, University of Southampton, Southampton, SO17 1BJ, UK}

\author{Eric Plum}
\email{erp@orc.soton.ac.uk}
\affiliation{Optoelectronics Research Centre, University of Southampton, Southampton, SO17 1BJ, UK}
\affiliation{Centre for Photonic Metamaterials, University of Southampton, Southampton, SO17 1BJ, UK}

\author{Yongmin Jung}
\affiliation{Optoelectronics Research Centre, University of Southampton, Southampton, SO17 1BJ, UK}

\author{Venkatram Nalla}
\affiliation{Centre for Disruptive Photonic Technologies, School of Physical and Mathematical Sciences and The Photonics Institute, Nanyang Technological University, Singapore 637371}

\author{Cosimo Lacava}
\affiliation{Optoelectronics Research Centre, University of Southampton, Southampton, SO17 1BJ, UK}

\author{Kevin F. MacDonald}
\affiliation{Optoelectronics Research Centre, University of Southampton, Southampton, SO17 1BJ, UK}
\affiliation{Centre for Photonic Metamaterials, University of Southampton, Southampton, SO17 1BJ, UK}

\author{Periklis Petropoulos}
\affiliation{Optoelectronics Research Centre, University of Southampton, Southampton, SO17 1BJ, UK}

\author{David J. Richardson}
\affiliation{Optoelectronics Research Centre, University of Southampton, Southampton, SO17 1BJ, UK}

\author{Nikolay I. Zheludev}
\email{niz@orc.soton.ac.uk}
\homepage{www.nanophotonics.org.uk}
\affiliation{Optoelectronics Research Centre, University of Southampton, Southampton, SO17 1BJ, UK}
\affiliation{Centre for Photonic Metamaterials, University of Southampton, Southampton, SO17 1BJ, UK}
\affiliation{Centre for Disruptive Photonic Technologies, School of Physical and Mathematical Sciences and The Photonics Institute, Nanyang Technological University, Singapore 637371}

\date{\today}

\maketitle

\textbf{
Recently, coherent control of the optical response of thin films of matter in standing waves has attracted considerable attention, ranging from applications in excitation-selective spectroscopy and nonlinear optics to demonstrations of all-optical image processing. Here we show that integration of metamaterial and optical fibre technologies allows the use of coherently controlled absorption in a fully fiberized and packaged switching metadevice. With this metadevice, that controls light with light in a nanoscale plasmonic metamaterial film on an optical fibre tip, we provide proof-of-principle demonstrations of logical functions XOR, NOT and AND that are performed within a coherent fully fiberized network at wavelengths between 1530~nm and 1565~nm. The metadevice performance has been tested with optical signals equivalent to a bitrate of up to 40~Gbit/s and sub-milliwatt power levels.
Since coherent absorption can operate at the single photon level and also with 100~THz bandwidth, we argue that the demonstrated all-optical switch concept has potential applications in coherent and quantum information networks.
}

\begin{figure} [tb]
\includegraphics[width=85mm]{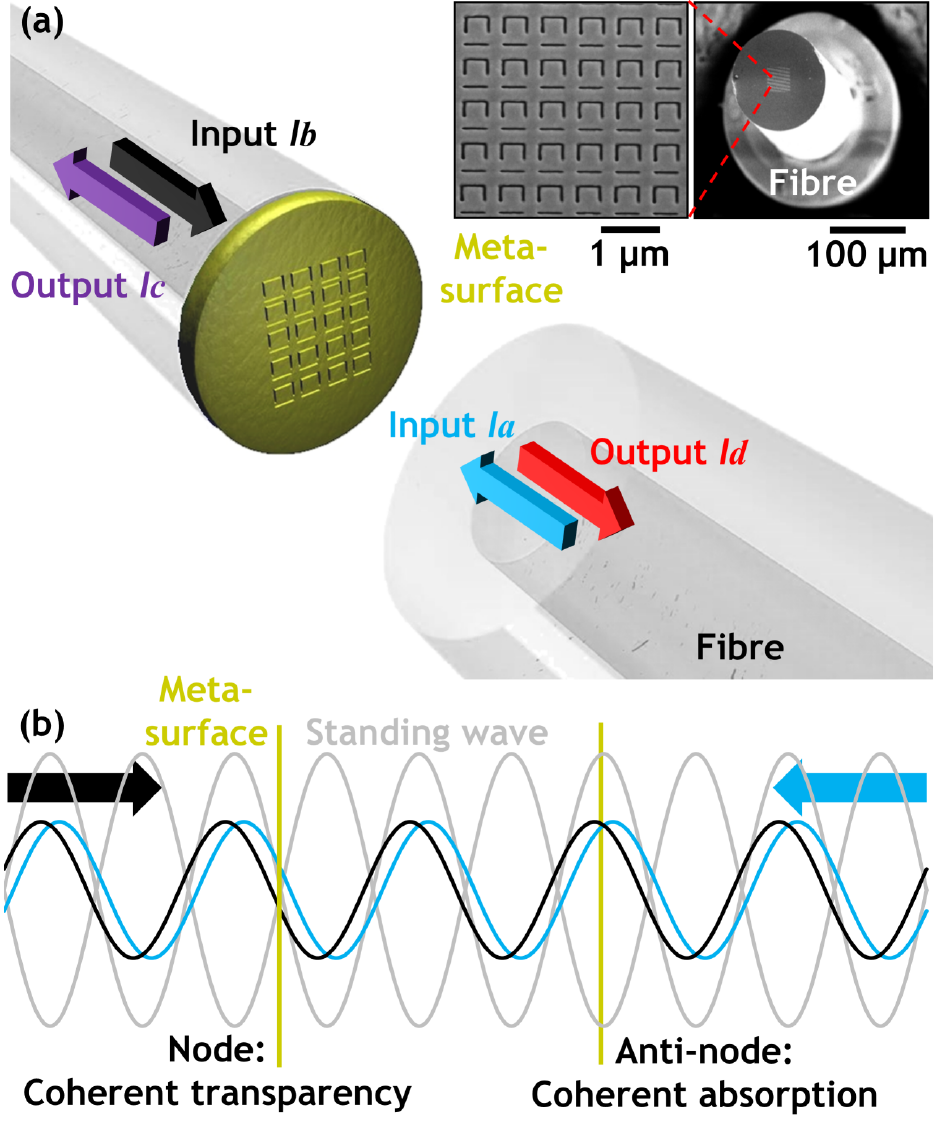}
\caption{\label{Fig_Concept}
\textbf{Coherent interaction of light with light on a metasurface. } (a)~Coherent optical input signals $I_a$ and $I_b$ interact on a metasurface absorber, generating output signals $I_c$ and $I_d$. The metasurface has been fabricated by nanostructuring the central $25 \times 25~\mu\text{m}^2$ of a 70-nm-thick gold layer covering the cleaved end-face of a polarization-maintaining single-mode telecommunications fibre shown by the insets (SEM images). (b)~The counterpropagating coherent input signals form a standing wave and --- depending on their phase difference --- the metasurface can be located at a position of destructive interference of electric fields (node) where absorption is suppressed or a position of constructive interference (anti-node) where absorption is increased. In the ideal case, absorption can be controlled from 0\% to 100\%.
}
\end{figure}

All-optical signal processing fundamentally relies on modulation of one optical signal with another. Therefore all-optical logical functions have long been perceived as the exclusive domain of nonlinear optics \cite{AllOpticalLogicGatesReview_2014}, which requires a minimum level of intensity to activate the nonlinear material response and faces trade-offs between magnitude and speed of the nonlinearity involved \cite{Boyd_NLO_3rdEd_p211, NonlinearSiChip_2004, SiResonators_2007, fJswitching_2010, OpticalSignalProcessing_2014}. However, recently it was shown that an effective nonlinear response may be derived from coherent interaction of light with light on linear materials of substantially sub-wavelength thickness \cite{Fang2015}. In contrast to conventional optical nonlinearities, the effect has been shown to allow intensity-independent control over absorption of light, from almost 0\% to almost 100\% \cite{Zhang2012}, with 100~THz bandwidth \cite{Fang2014, Nalla2017} and even for single photon signals \cite{Roger2015}. The concept has enabled all-optical control of luminescence \cite{CoherentLuminescence2015, CoherentLuminescence2016}, redirection of light \cite{Shi2014, CoherentDeflection2017} as well as nonlinear \cite{CoherentNonlinear2015}, polarization \cite{CoherentPolarization2015} and quantum \cite{CoherentN00N2016, CoherentEntangled2017} effects in films of nanoscale thickness and excitation-selective spectroscopy \cite{CoherentSpectroscopy2017}. In particular, it has been predicted that coherent interaction of light on lossy ultrathin films could perform signal processing functions \cite{Fang2015} and proof-of-concept experiments in the static regime have been reported \cite{Papaioannou2016, Papaioannou2016APLphotonics}.
Here we report the realization of the first fully packaged, fiberized metamaterial device based on this concept. The metadevice uses a plasmonic nanostructure of substantially subwavelength thickness as a switchable absorber that allows the absorption of one optical pulse to be controlled by another coherent optical pulse. We demonstrate nonlinear input-output characteristics and all-optical operations analogous to logical functions NOT, AND and XOR at both kHz and GHz bitrates in a fiberized configuration assembled from standard telecoms components at wavelengths between 1530~nm and 1565~nm.

\section*{\large{Results}}

\begin{figure*} [tb]
\includegraphics[width=140mm]{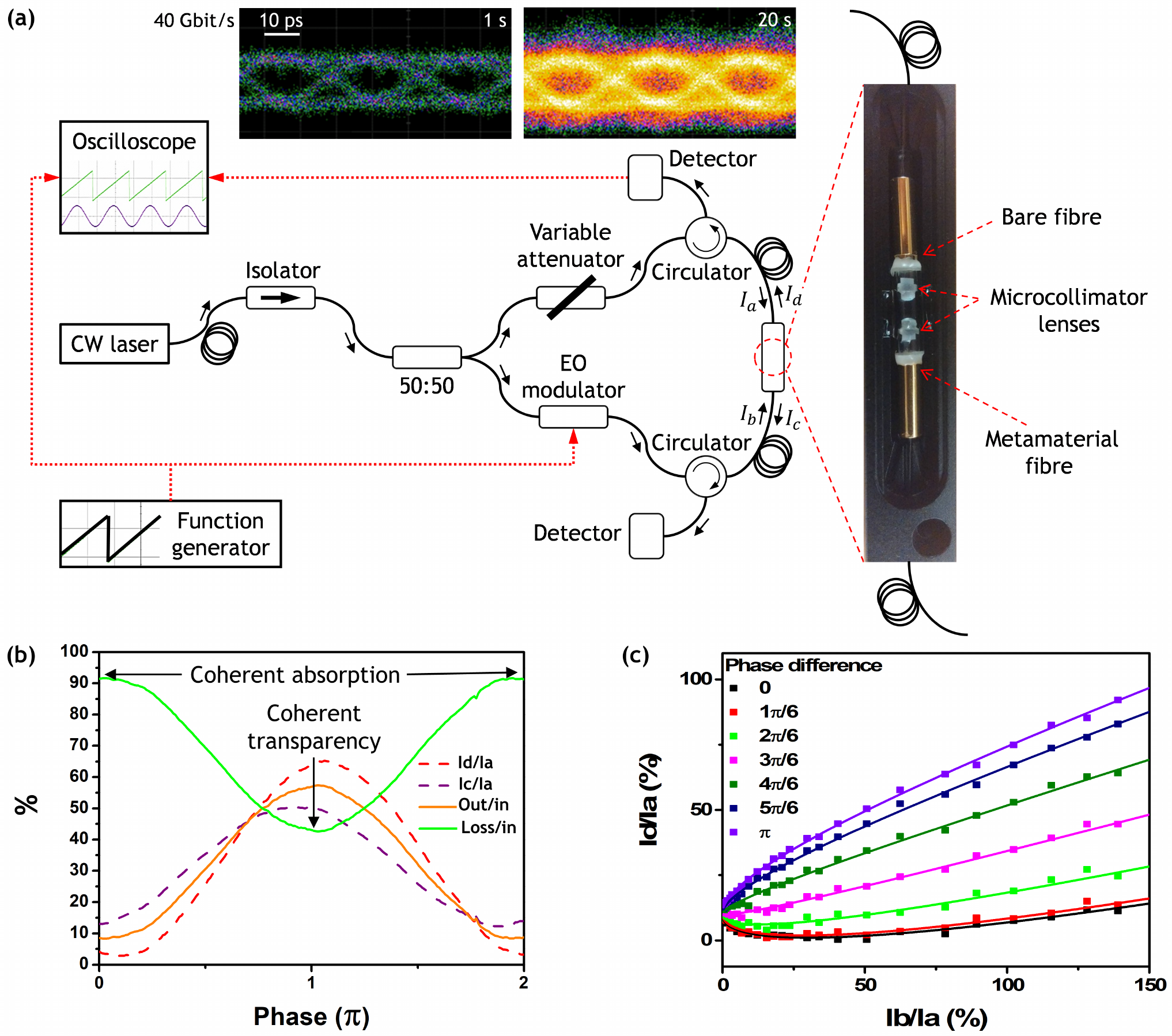}
\caption{\label{Fig_Setup}
\textbf{The packaged metadevice and its properties.} (a)~Schematic representation of the fully-fiberized experimental setup with a photograph of the packaged metadevice (without lid) consisting of the metasurface-covered fibre (Fig.~\ref{Fig_Concept}a) coupled to a bare fibre end using a pair of microcollimator lenses. The inset shows eye diagrams of output channel $I_d$ recorded for intensity modulation of input channel $I_b$ at 40~Gbit/s, where color indicates counts. (b)~Measured output intensities $I_c$ and $I_d$ (relative to $I_a$) as well as the total output power and metadevice losses (relative to the total input power) as a function of the phase difference between the input signals at the metasurface at 1550~nm wavelength. (c)~Measured output intensity $I_d$ (data points) relative to the fixed input intensity $I_a$ as a function of input intensity $I_b$ for various phase differences between the input beams, with fits (lines), at 1550~nm wavelength.
}
\end{figure*}

\begin{figure*} [tb]
\includegraphics[width=170mm]{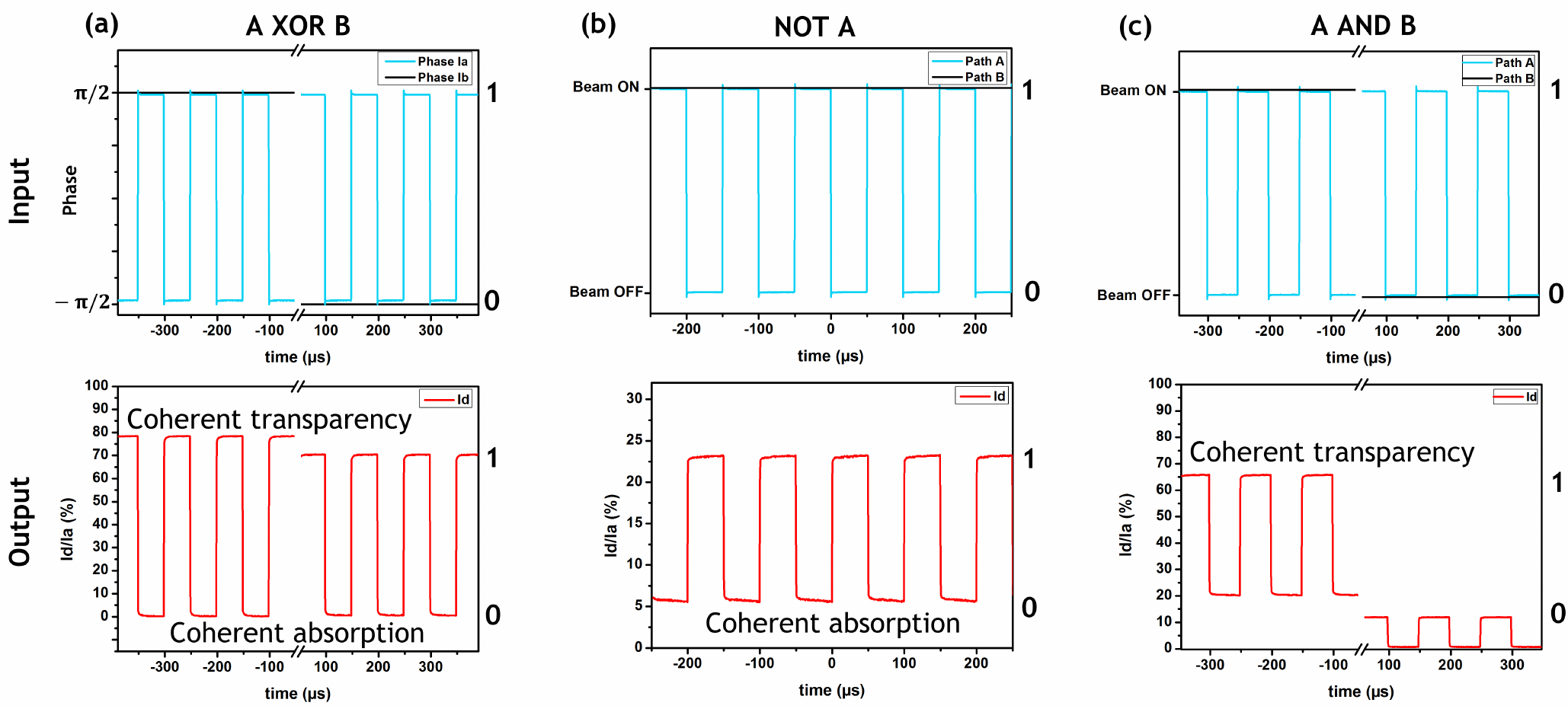}
\caption{\label{Fig_kHz}
\textbf{All-optical signal processing at kHz frequencies} at 1550~nm wavelength. (a)~XOR function between phase-modulated signals A and B producing an intensity-modulated output based on coherent absorption of identical bits and coherent transparency for opposite bits. (b)~NOT function on a single intensity-modulated signal A. The inversion of signal A in the presence of beam B (which is always on) results from coherent absorption of incoming signal pulses when the metasurface is located at a standing wave anti-node. (c)~AND function between intensity-modulated signals A and B resulting from coherent transparency of the metasurface for simultaneous illumination from both sides when the metasurface is located at a standing wave node. The logical states ``1" and ``0" are indicated on the right-hand side of each graph. The modulation frequency is 10~kHz in all cases. Minor signal distortions are due to the limited bandwidth of the waveform generator.
}
\end{figure*}

As illustrated in Fig.~\ref{Fig_Concept}a, the metadevice is based on controlling absorption of light with light on an ultrathin metamaterial absorber. It has two bidirectional ports, i.e.~two inputs, $I_a$ and $I_b$, and two outputs, $I_c$ and $I_d$. The input waves propagate in opposite directions and, provided that they are mutually coherent, co-polarized and with equal intensity, they will form a standing wave with electric field nodes and anti-nodes (Fig.~\ref{Fig_Concept}b). A sufficiently thin film may thus be placed at a node, where the electric field is zero due to destructive interference, or at an anti-node where the electric field amplitude is enhanced by constructive interference. Since truly planar structures interact with normally incident waves only via the electric field \cite{PlanarMM_BookChapter_2011}, this implies that a planar thin film placed at a node will be perfectly transparent, while the same thin film will be strongly excited if placed at an anti-node. With respect to absorption, which is limited to 50\% in planar materials illuminated by a travelling plane wave \cite{AbsorptionLimit_Abajo_2012}, standing wave illumination allows absorption to be controlled from 0\% to 100\% in the ideal case \cite{Fang2015}. Such ideal performance in the optical part of the spectrum that is relevant to optical fibre technology may be approximated with materials that are thin compared to the optical wavelength and that exhibit equal transmission and reflection in addition to close to 50\% travelling wave absorption, for example nanostructured plasmonic metamaterials \cite{Zhang2012} and 30-layer graphene \cite{CoherentGraphene_2014}.

The planar absorber used here is a plasmonic metamaterial consisting of a 70-nm-thick gold film perforated with an array of asymmetrically split ring apertures. This metamaterial design has been chosen since it performed well in free-space demonstrations of coherent light absorption and transparency \cite{Zhang2012} and as the well-known dependence of its transmission, reflection and absorption characteristics on aperture size and geometry \cite{Combinatorial_ASR_2011} allows for easy optimization throughout the optical telecommunications bands. Our metadevice operates at wavelengths around $\lambda=1550$~nm, therefore its 70~nm thickness corresponds to $\lambda/22$. The metamaterial structure has been fabricated by thermal evaporation
of gold and subsequent focused ion beam milling
on a $25 \times 25~\mu\text{m}^2$ area covering the core of a cleaved polarization-maintaining telecommunications fibre, see inset of Fig.~\ref{Fig_Concept}a. The fibre output was coupled to a second cleaved optical fibre using two microcollimator lenses to realize the first in-line fibre metadevice to our knowledge, see Fig.~\ref{Fig_Setup}a. Care was taken to achieve alignment of the symmetry axis of the metamaterial with the slow axis of the polarization maintaining fibres, see Methods for details.

The metadevice has standard FC/APC fibre connectors and it was characterized in a fibre interferometer assembled from standard polarization maintaining fibre components, see Fig.~\ref{Fig_Setup}a. The output of a fibre-coupled CW laser
was split along two paths of similar length, with one path containing an electro-optical phase or intensity modulator and the other containing a variable attenuator to allow balancing of the power propagating along the two paths. The paths were then recombined within the metadevice and the output signals were detected via circulators using an oscilloscope, see Methods for a more detailed description. We note that practical applications would require some means of active stabilisation of the optical path lengths, however, the system shown here was sufficient for characterising the principle of operation of the fiberized metadevice. This is illustrated by the eye diagrams in Fig.~\ref{Fig_Setup}a, which show that the ``eye" closes on a timescale of seconds due to phase drift in the interferometer that is used to characterize the metadevice.

Fig.~\ref{Fig_Setup}b shows the phase-dependent output intensities $I_c$ and $I_d$ relative to the input intensity $I_a=I_b$ as a function of the phase difference between the inputs at 1550~nm wavelength. The overall output intensity can be controlled from about 9\% to about 57\% of the total intensity entering the metadevice, where a low output intensity corresponds to constructive interference of the incident waves on the metasurface and thus coherent absorption, while a high output intensity corresponds to coherent transparency. Both output signals display a similar phase-dependence, however, the phase-dependent output intensity $I_d$ offers somewhat higher contrast and therefore we focus on this output. We note that, for an ideal metadevice containing a perfectly symmetric ultrathin absorber with 50\% single beam absorption and no other loss mechanisms, both output intensities would be identical and modulated from complete absorption to perfect transmission. Differences between the measured output channels arise from the asymmetric construction of our metadevice that contains a metasurface fabricated on the glass/air interface of one of the optical fibres. The whole metadevice exhibits about 24\% single beam transmission, 18\% (8\%) reflection and 58\% (68\%) losses for a single input signal $I_a$ ($I_b$). However, only part of these losses correspond to absorption in the metasurface that can be coherently controlled, while other sources of losses include the fibre connections of the metadevice, scattering and unwanted reflections within the microcollimator and fabrication imperfections such as imperfect alignment of the metasurface orientation with the slow axis of the fibres.

The nonlinear functionality of the linear metadevice is illustrated by Fig.~\ref{Fig_Setup}c, which shows how the intensity of output $I_d$ depends on the intensity of input $I_b$ while $I_a$ remains constant for various phase differences between the input beams. The measured output intensity $I_d$ as a function of input $I_b$ is nonlinear and generally follows the behaviour predicted by Fang et al \cite{Fang2015}. For input phase differences of less than $\pi/2$ it is also nonmonotonic --- counterintuitively the output intensity decreases with increasing input intensity and reaches a minimum before increasing when $I_b$ becomes large.
For an input phase difference of $\pi/2$, changes in the measured output intensity $I_d$ are approximately proportional to changes in the input intensity $I_b$. For larger input phase differences $I_d$ as a function of $I_b$ flattens with increasing $I_b$, but is steep for small $I_b$ suggesting possible applications in small signal amplification. Thus, the results presented in Fig.~\ref{Fig_Setup}b,c show that large changes of the metadevice output result from modulation of phase or intensity of one of the metadevice inputs.

All-optical signal processing operations with input/output relations analogous to logical functions may now be realized by exploiting coherent transparency and/or coherent absorption in the metadevice. For example, consider mutually coherent, binary, phase-modulated signals, where the logical states, ``+" and ``-", are represented by equal intensity but opposite phase. Constructive interference of identical bits A and B from both input signals will lead to coherent absorption on the metasurface, while destructive interference of opposite bits will lead to coherent transparency, corresponding to an intensity-modulated output A XOR B. The behavior of an ideal metadevice is illustrated in Table~\ref{my-table-phase} and the measured behaviour of the experimental metadevice is shown in Fig.~\ref{Fig_kHz}a for a modulation frequency of 10~kHz. The measurements clearly demonstrate an XOR function with large contrast ($>10\times$) between the output states. 
The intensity of the logical ``1" is about 30\% lower than in the ideal case due to losses within the metadevice. We note that the XOR function can be inverted to A XNOR B by providing an additional external phase shift $\theta_\text{ext}=\pi$ to one of the input signals. Furthermore, a fixed input signal B of + or - could be used to map the phase-modulated signal A to an intensity-modulated signal with (NOT A) or without (IDENTITY A) inversion.

\begin{table}[]
\centering
\caption{Logical functions between mutually coherent, equal intensity, phase-modulated bits A and B ($I_a=I_b=1$)}
\label{my-table-phase}
\begin{tabular}{cc|c|c}
\multicolumn{2}{c|}{Input states} & \multicolumn{2}{c}{Ideal output intensities $I_c=I_d$} \\
         &           & $\theta_\text{ext}=0$ & $\theta_\text{ext}=\pi$ \\
A        & B         & A XOR B & A XNOR B  \\ \hline
+        & +         & 0       & 1         \\
+        & -         & 1       & 0         \\
-        & +         & 1       & 0         \\
-        & -         & 0       & 1
\end{tabular}
\end{table}

The metadevice can also perform signal processing operations analogous to logical functions on intensity-modulated optical data. The simplest example is a NOT function. Such an inversion of an intensity-modulated signal A is achieved by leaving input beam B always switched on with its phase adjusted such that coherent absorption will occur for simultaneous illumination of the metasurface from both sides ($\theta_\text{ext}=0$). Input pulses A (logical 1) will be coherently absorbed resulting in low output (logical 0). On the other hand, low input signals A (logical 0) will allow some light from input B to reach the outputs (logical 1). For an ideal metadevice, the expected output intensities are 0\% and 25\% of the input intensity, respectively, and our experimental device achieves about 5\% and 23\% at a modulation frequency of 10~kHz, which is more than sufficient for distinguishing the logical states, see Fig.~\ref{Fig_kHz}b.

While the NOT function was based on coherent absorption, an AND function between binary intensity-modulated signals can be realized by exploiting coherent transparency. In this case, a phase shift is applied to one input signal such that simultaneous illumination of the metasurface from both sides leads to coherent transparency ($\theta_\text{ext}=\pi$). For an ideal device, this would lead to 100\% output intensity for interaction of two pulses on the metasurface and at least $4\times$ lower output intensity for any other combination of input bits, see table~\ref{my-table-amplitude}. Experimentally we observe the AND function with more than $3\times$ contrast between the logical output states at a modulation frequency of 10~kHz. In principle, other logical functions including XOR and OR for intensity-modulated signals can be realized for suitable choices of $\theta_\text{ext}$ \cite{Papaioannou2016APLphotonics}.

\begin{table}[]
\centering
\caption{Logical function A AND B between mutually coherent, intensity-modulated bits A and B}
\label{my-table-amplitude}
\begin{tabular}{cc|c}
\multicolumn{2}{c|}{Input states} & \multicolumn{1}{c}{Ideal output intensities $I_c=I_d$} \\
         &           & $\theta_\text{ext}=\pi$ \\
A=$I_a$  & B=$I_b$   & A AND B \\ \hline
1        & 1         & 1       \\
1        & 0         & 0.25    \\
0        & 1         & 0.25    \\
0        & 0         & 0
\end{tabular}
\end{table}

\begin{figure*} [tb]
\includegraphics[width=170mm]{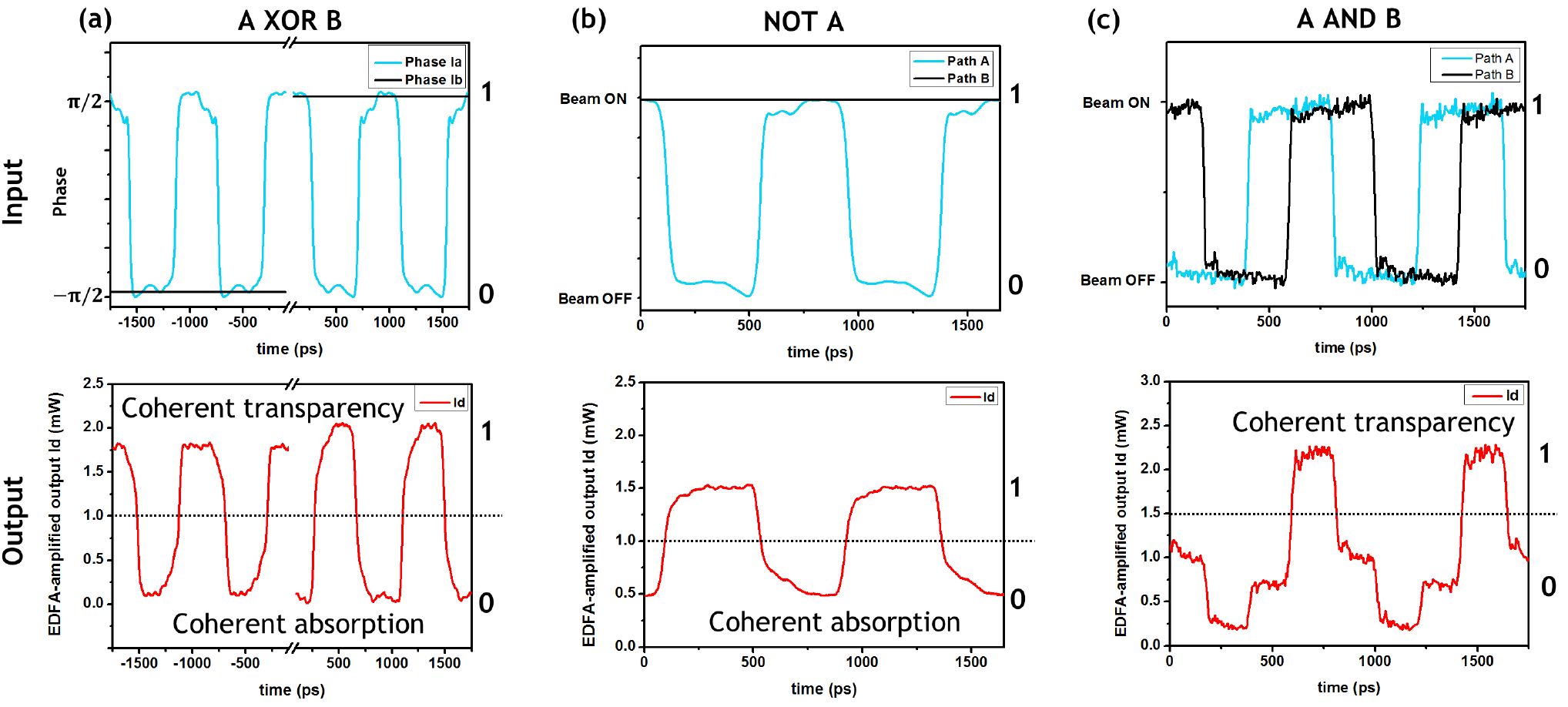}
\caption{\label{Fig_GHz}
\textbf{All-optical signal processing at 1.2~GHz} and 1550~nm wavelength. (a)~XOR function between phase-modulated signals A and B producing an intensity-modulated output based on coherent absorption being turned on by identical bits and turned off by opposite bits. (b)~NOT function on a single intensity-modulated signal A in the presence of a constant beam B that causes coherent absorption of incoming signal pulses. (c)~AND function on two intensity-modulated signals A and B resulting from coherent transparency of the metasurface for simultaneous illumination from both sides when the metasurface is located at a standing wave node. The different noise level in panel (c) is due to a different experimental configuration, see Methods. The logical states ``1" and ``0" are indicated on the right-hand side of each graph.
}
\end{figure*}

\begin{figure} [tb]
\includegraphics[width=65mm]{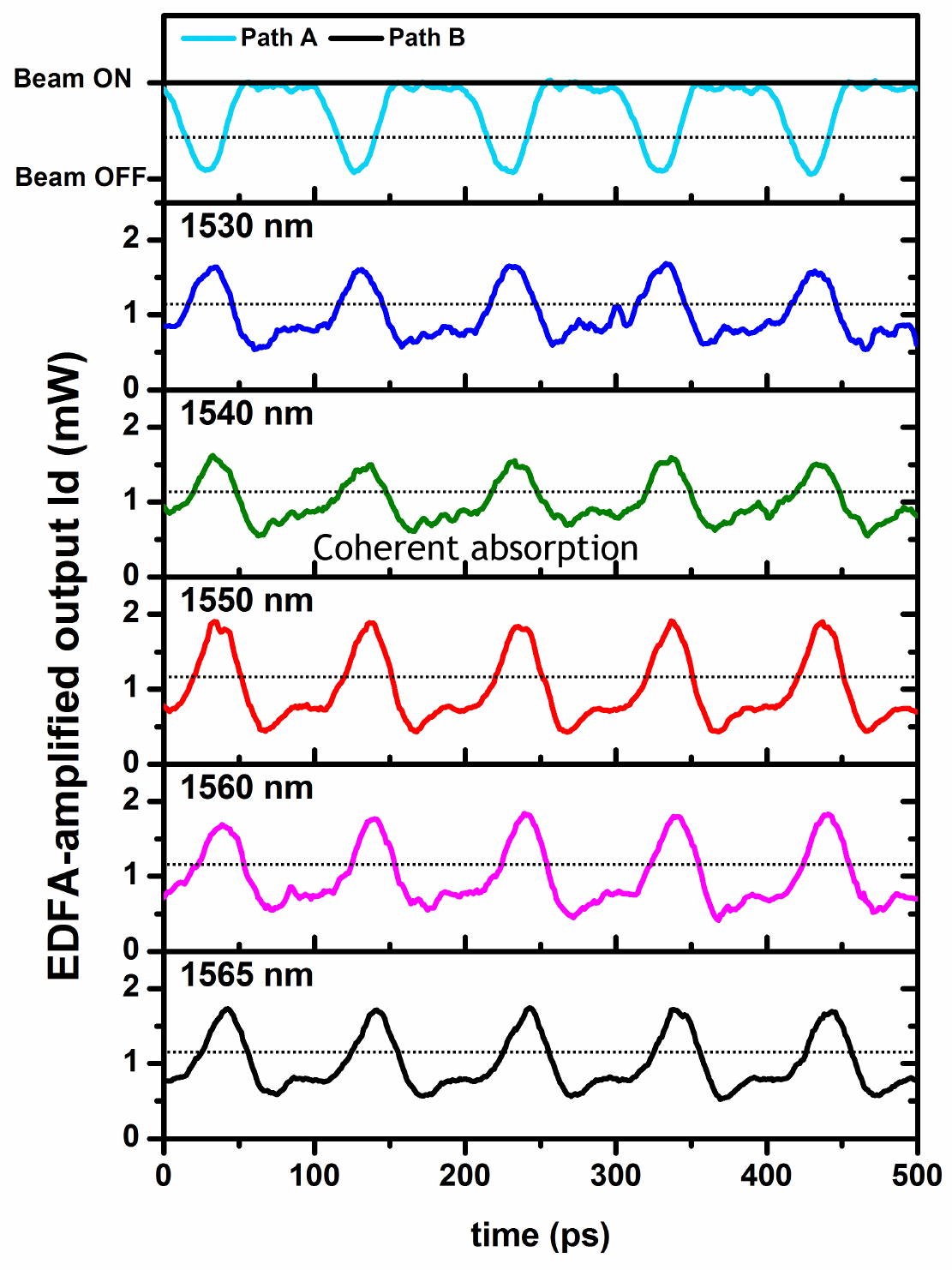}
\caption{\label{Fig_40Gbit}
\textbf{Broadband inversion NOT A of a 40~Gbit/s signal A} at wavelengths from 1530~nm to 1565~nm. The signal corresponds to an intensity-modulated bit pattern 1011 repeating at 10~GHz (top) and the inverted output signal was measured at different wavelengths (bottom). Beam B has a constant intensity and causes coherent absorption of incoming signal pulses. The logical states ``1" and ``0" are separated by a horizontal dotted line on each graph.
}
\end{figure}

In practical systems, optical signals transferred by optical fibres are modulated at GHz frequencies rather than kHz, and also make use of a range of optical wavelengths. Therefore we tested the metadevice also at 5-6 orders of magnitude higher modulation rates and over wavelengths from 1530~nm to 1565~nm as illustrated in Fig.~\ref{Fig_GHz} and Fig.~\ref{Fig_40Gbit}. The output of the fibre interferometer was amplified with an erbium-doped fibre amplifier (EDFA) which provided an average output power of 1~mW. This conveniently ensures that the threshold power between logical 0 and logical 1 will always be close to 1~mW. Fig.~\ref{Fig_GHz}a shows the XOR function on phase-modulated signals (as described above) now at a modulation frequency of 1.2~GHz, while Fig.~\ref{Fig_GHz}b shows the NOT function on an intensity-modulated signal at 1.2~GHz. An AND function was realized by intensity modulation of the laser light entering the interferometer (i.e.~modulation before the 50:50 splitter shown in Fig.~\ref{Fig_Setup}a) and introducing a path difference in the interferometer arms in order to delay the modulated signals A and B relative to each other as illustrated by Fig.~\ref{Fig_GHz}c. This is equivalent to an intensity modulated bit sequence 0011 in channel A and 1001 in channel B, resulting in an output 0001, i.e.~A AND B, in the detected output channel.

Fig.~\ref{Fig_40Gbit} shows the NOT function of the metadevice at a frequency of 10~GHz. The periodic input signal A was generated with a bit pattern generator and is equivalent to a bit sequence of 1011 that is inverted to become 0100 at a bitrate of 40~Gbit/s. Measurements at wavelengths from 1530~nm up to 1565~nm wavelength show successful signal inversion illustrating the broadband nature of the underlying effect of coherent absorption across the wavelength range accessible to the tuneable laser used in the experiment. Considering 50~$\mu$W peak power in each input channel and 25~ps per bit, the modulator's energy consumption is 2.5~fJ per bit under our experimental conditions, which corresponds to about 20 000 photons per bit. Given that coherent absorption of single photons has been demonstrated \cite{Roger2015}, we expect that energy consumption in the aJ per bit regime should be possible with a sufficiently sensitive detection system.

The modulation results obtained at GHz frequencies are similar to those obtained at kHz frequencies. At GHz frequencies, there is some signal distortion apparent in both the signals used to drive the phase and intensity modulators as well as the measured modulation of the output signal, which has slightly smaller contrast. These distortions arise from the frequency response and background noise of the modulators and amplifiers used in the experiments.
While our experimental equipment does not allow us to test the performance of the modulator in the fibre environment beyond 40~Gbit/s, we expect that the metadevice can in principle operate at much higher frequencies. Indeed, the underlying phenomena of coherent absorption and coherent transparency in plasmonic metamaterials occur on timescales as short as 10~fs implying a potential bandwidth on the order of 100~THz \cite{Nalla2017}.
Comparison of the absorption characteristics of such an ultrafast coherent absorber to those of our metadevice indicates that the metamaterial used in our metadevice may be expected to efficiently absorb pulses as short as 40~fs, corresponding to a potential bandwidth of 10s of THz for our device, see Supplementary Information. We admit, however, that such bandwidth will be difficult to realize in a fiberized device due to dispersion limitations of the fibres.

\section*{\large{Discussion}}

Even though we report a proof-of-principle demonstration, the metadevice shows that coherent interaction of light with light on ultrathin films can be used to perform signal processing functions with high bandwidth and high contrast on signals carried by telecoms fibres. Further improved device performance would result from a more symmetric design where the absorber of nanoscale thickness is in contact with the same material on both sides, for instance by developing appropriate splicing techniques. The absorber could also be placed in between two optical fibre ends spaced by a nanoscale distance and surrounded by index matching liquid, which would also help eliminate incoherent scattering and absorption losses. In addition, the metamaterial design could be further improved to achieve 50\% single beam absorption and identical 25\% transmission and reflection characteristics from both sides as well as polarization independence of the nanostructure. The possibility of replacing the metamaterial with multi-layer graphene could also be explored.

All-optical signal processing applications based on controlling absorption of light with coherent light promise extremely high bandwidth and extremely low energy requirements, but they will require mutually coherent signals and phase stability. Mutually coherent signals are most easily achieved in local systems, where multiple signals are driven by the same seed laser and as locally coherent networks are becoming part of the mainstream telecommunications agenda \cite{CoherentCommunications2010, Slavic_AmplitudePhaseRegenerator_2010, CoherentCommunications2014}, coherent all-optical data processing may become a realistic proposition, particularly in the miniaturized optical chips of integrated optics and silicon photonics \cite{CoherentCommunicationsSiPhotonics2014}, where phase stability is more easily achieved than in large-scale fibre networks. In order to go beyond simple single-step logical functions, cascading of multiple coherent all-optical logical operations will also need to be explored. Such cascading is likely to require signal generation techniques as the XOR function converts phase-shift keying to amplitude-shift keying (in principle without insertion loss and with unlimited contrast), while the AND function suffers from limited contrast of 6~dB (in principle without insertion loss) and the NOT function suffers from insertion loss of 6~dB (in principle with unlimited contrast). Beyond all-optical data processing, potential applications of coherent metadevices include small-signal amplification and coherence filtering \cite{Zhang2012}.

The metadevice demonstrated here is an example of the broad range of opportunities arising from fibre-integration of metasurfaces that could also be used to control, for example, focusing, polarization, spectral characteristics, propagation direction and angular momentum of light \cite{MetasurfaceFibres2013, MetasurfaceFibres2015, MetamaterialPhaseGradientMS_2017}.

In summary, we report the first example of a metamaterial-based device for all-optical signal processing that is compatible with optical telecommunications fibre components. The multi-functional metadevice can perform effectively nonlinear signal processing functions including input/output relations analogous to XOR, AND and NOT operations and the underlying mechanism of coherent transparency and coherent absorption is compatible with single photon signals and 100~THz bandwidth. We therefore anticipate that such metadevices can provide solutions for quantum information networks as well as orders-of-magnitude improvements in speed and energy consumption over existing nonlinear approaches to all-optical signal processing in coherent information networks.

\section*{\large{Methods}}

\subsection*{Metadevice fabrication}

For the realization of the metadevice, the metamaterial was fabricated on the cleaved end face of a polarization-maintaining single-mode fibre as described above. The metamaterial nanostructure was fabricated with its symmetry axis aligned with the slow axis of the Panda style fibre. In order to allow metamaterial illumination from either side, the metamaterial-covered fibre was interfaced with a second polarization maintaining fibre using a microcollimator arrangement consisting of a pair of microlenses.
More specifically, the metamaterial-covered fibre end and an anti-reflection-coated microlens were held in place by a fibre ferrule, leaving a gap of approximately 10~$\mu$m between the two components.
The second fibre also had its end cleaved and was mounted in the same way. Using kinematic mounts, both fibre ferrules were aligned to each other by maximizing the light coupling efficiency and polarization contrast. The aligned microcollimator arrangement was fixed with UV-curable glue by UV exposure, stabilized with an outer glass shell and placed in a plastic housing for extra protection, see Fig.~\ref{Fig_Setup}a.

\subsection*{Experimental metadevice characterization}

The experimental characterization including modulation of absorption and total output intensity of light with light, effective nonlinearity and all-optical logical functions took place in the above-mentioned setup (see Fig.~\ref{Fig_Setup}a). All fibre components were based on Panda style polarization-maintaining single-mode fibres. In all optical experiments, the incident electric field is oriented parallel to the symmetry axis of the metamaterial nanostructure (slow axis of the fibre). The transmission losses of the metadevice including encapsulation losses and fibre splice losses are about 5.2~dB.

Measurements recorded at kHz frequencies used the 180~$\mu$W output of a fibre-coupled 1550~nm wavelength CW laser diode and were recorded using InGaAs photodetectors and an oscilloscope (Agilent Technologies DSO7104A). They have been calibrated by taking into account the insertion losses of the commercial fibre-coupled components, so that $I_a$, $I_b$, $I_c$ and $I_d$ in Fig.~\ref{Fig_Setup}b,c and Fig.~\ref{Fig_kHz} correspond to the intensities entering and leaving the metadevice's fibre connectors. The peak input power at the metadevice input connector $I_a=I_b$ was 10~$\mu$W. The modulators used were low-loss electro-optical 10~Gbit/s phase or intensity modulators (EOspace) driven by a waveform generator (AM300 by Rohde \& Schwarz).

Measurements recorded at GHz frequencies used a fibre-coupled tuneable CW laser (ID Photonics CoBrite-DX4) and show the EDFA-amplified output power detected by an oscilloscope (Agilent Infiniium DCA-J 86100C) as described in the main text. The peak input power at the metadevice input connector $I_a=I_b$ is about 100~$\mu$W (Fig.~\ref{Fig_GHz}a,b), 30~$\mu$W (Fig.~\ref{Fig_GHz}c) and 50~$\mu$W (Fig.~\ref{Fig_40Gbit}). The modulators used were low-loss electro-optical 10~Gbit/s phase or intensity modulators (EOspace) driven by an arbitrary waveform generator (Tektronix AWG7122C) and a radio frequency amplifier (LA Techniques) in case of Fig.~\ref{Fig_GHz}a,b, while a bit pattern generator (SHF 12100 B) was used in case of Fig.~\ref{Fig_Setup}a, Fig.~\ref{Fig_GHz}c and Fig.~\ref{Fig_40Gbit}.

\subsection*{Data availability}

Following a period of embargo, the data from this paper will be available from the University of Southampton ePrints research repository: http://doi.org/10.5258/SOTON/D0172

\section*{\large{Acknowledgements}}

The authors thank Daniele Faccio, Jun-Yu Ou and Vassili Savinov for advice and fruitful discussions.
This work is supported by the UK's Engineering and Physical Sciences Research Council (grant EP/M009122/1) and the MOE Singapore (grant MOE2011-T3-1-005).

\section*{\large{Supplementary Information}}

\begin{figure} [t]
\includegraphics[width=78mm]{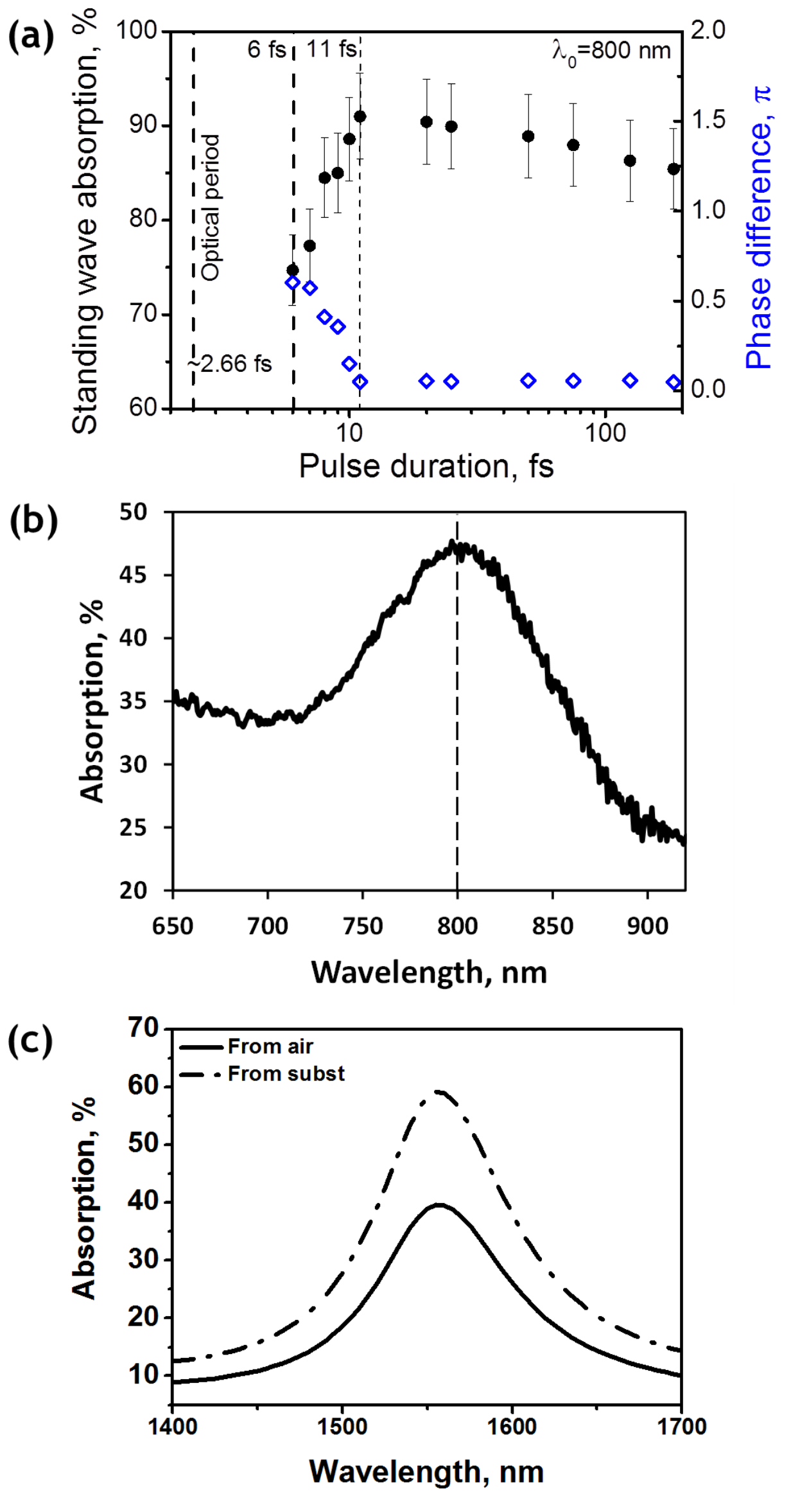}
\caption{\label{Fig_fs}
\textbf{Interaction of femtosecond pulses with a thin absorber.} (a)~Coherent absorption of counterpropagating femtosecond pulses measured on a freestanding plasmonic metamaterial for different pulse durations at 800~nm wavelength. The phase difference between the output beams (diamonds) becomes dependent on the pulse duration as absorption (circles) deteriorates for pulses shorter than 11~fs. (b)~Measured absorption spectrum of the plasmonic metamaterial used in panel (a) for illumination by a single beam of light. (c)~Simulated absorption spectrum of the metamaterial used in the metadevice of the main manuscript for illumination from outside the fibre (solid, $I_a$) and within the fibre (dashed, $I_b$).
}
\end{figure}

While our experimental equipment does not allow us to test the performance of the main manuscript's fiberized modulator beyond 40~Gbit/s, we expect that the metadevice can in principle operate at much higher frequencies. Indeed, it was recently shown that the underlying phenomena of coherent absorption and coherent transparency in plasmonic metamaterials occur on timescales as short as 10~fs implying a potential bandwidth on the order of 100~THz \cite{Nalla2017}. 
We demonstrate this by measuring coherent absorption of femtosecond pulses in a freestanding plasmonic metamaterial that is similar to the one employed in our metadevice, see Fig.~\ref{Fig_fs}a. The metamaterial consists of a freestanding gold film of 60~nm thickness perforated with a split ring aperture array of a smaller 320~nm period and it exhibits an absorption peak at the experimental wavelength of 800~nm that was determined by the available ultra-short pulse laser. Our measurements show that coherent absorption starts to drop for pulses shorter than 11~fs. We note that a slow decline of coherent absorption upon increase of pulse duration from 11~fs to 185~fs is explainable by the onset of nonlinear absorption at high fluences. The duration of the shortest optical pulse that can be efficiently absorbed is linked to the plasmon relaxation time in gold and the spectral width of the plasmonic response (Fig.~\ref{Fig_fs}b) as interaction of spectral components of pulses outside of the plasmonic absorption line will not lead to efficient coherent absorption. The absorption resonance of our metadevice has roughly the same spectral width in terms of wavelength, but operates at about twice the wavelength. Assuming the same time bandwidth product in both cases, this implies that our metadevice will efficiently absorb pulses as short as 40~fs, corresponding to a potential bandwidth of 10s of THz for our device (Fig.~\ref{Fig_fs}c). We admit, however, that such bandwidth will be difficult to realize in a fiberized device due to dispersion limitations of the fibres.\\~\\

\section*{\large{Supplementary Methods}}

\subsection*{Coherent control of femtosecond pulses}

The metamaterial absorber for experiments with femtosecond pulses (Fig.~\ref{Fig_fs}a,b) is a nanostructured free-standing gold film of 60~nm thickness. It was fabricated by thermal evaporation of gold on a 50-nm-thick silicon nitride membrane, followed by silicon nitride removal by reactive ion etching and nanostructuring of the remaining free-standing gold film by gallium focused ion beam milling. The gold film is perforated with an array of $320\times320~$nm$^2$ split ring apertures that has an overall size of $50\times50~\mu$m$^2$ and a resonant absorption peak around 800~nm wavelength.

Coherent absorption was measured as a function of pulse duration using a free-space setup and a 6~fs mode-locked Ti:sapphire laser (Femtolasers Rainbow) operating at a central wavelength of 800~nm and equipped with a pulse shaper (Biophotonics MIIPS). This was performed by splitting the output of the pulse shaper along two paths of identical length, which are recombined on the metamaterial absorber such that constructive interference occurs on the metamaterial and the light that remained after interaction with the nanostructure was detected.

\subsection*{Metadevice simulations}

The absorption spectra of the metamaterial used in the metadevice of the main manuscript (Fig.~\ref{Fig_fs}c) were modelled by simulating a single metamaterial unit cell with periodic boundary conditions and normal incidence illumination using finite element modelling (COMSOL Multiphysics 3.5a) in three dimensions. The permittivity of gold was taken from ref.~\onlinecite{Palik} and the permittivity of glass was assumed to be 2.0736.


\end{document}